\documentclass{ws-p8-50x6-00}

\begin{document}

\title{Nuclear Aspects of Nucleosynthesis in Massive Stars}

\author{T. Rauscher$^{1,2}$, R.D. Hoffman$^3$, A. Heger$^2$, S.E. Woosley$^2$}

\address{$^1$ Departement f\"ur Physik und Astronomie, Universit\"at Basel,
CH-4056 Basel, Switzerland
\\E-mail: Thomas.Rauscher@unibas.ch}

\address{$^2$ Department of Astronomy and Astrophysics, University of
California at Santa Cruz, Santa Cruz, CA 95064, USA}

\address{$^3$ Nuclear Theory and Modeling Group, L-414, Lawrence Livermore
National Laboratory, Livermore, CA 94551, USA}


\maketitle

\abstracts{
Preliminary results of a new set of stellar evolution and 
nucleosynthesis calculations for
massive stars are presented.
These results were obtained with an extended reaction network up to
Bi. The discussion focuses on the importance of nuclear
rates in pre- and post-explosive nucleosynthesis.
The need for further experiments to study 
optical $\alpha$+nucleus potentials
is emphasized.}

\section{Introduction}
Nuclear reactions play a major role not only in the nucleosynthetic processes
determining the elemental abundances in the solar system and the Galaxy but
also for determining structure and final fate of a star.
Massive stars ($>8 M_{\odot}$) experience a number of burning phases before
they explode as type II supernovae after the collapse of the Fe core.
Important nuclear reactions in the late burning stages and in the explosion
proceed on isotopes experimentally not sufficiently well investigated or on
unstable nuclei which cannot be studied in the laboratory. Thus, astrophysics
requires a sound theoretical understanding of nuclear reactions and tests
our knowledge at the extremes.

Numerous studies have been devoted to the evolution of
such stars and their nucleosynthetic yields. However, our knowledge of both
the input data and the physical processes affecting the development of
these objects has improved dramatically in recent years. Thus, it became
worthwhile to attempt to improve on and considerably extend the previous
investigations on pre-- and post--collapse evolution and nucleosynthesis.
Here we present first results for a 15 $M_{\odot}$ stellar model with improved
stellar and nuclear physics. In this report we mainly concentrate on a few
of the
nuclear physics issues involved, a more extended report including all details
of the simulation will be published elsewhere\cite{hhrw00,rhhw00}.
Below, we discuss the prediction of nuclear rates in the statistical model
and, specifically, the treatment of $\alpha$-particle capture on isospin
conjugated targets. The importance of obtaining more information on 
$\alpha$+nucleus potentials is emphasized separately.

\section{Nuclear Reactions}
\begin{figure}[t]
\begin{center}
\includegraphics*[bb=50 55 792 612,height=16pc]{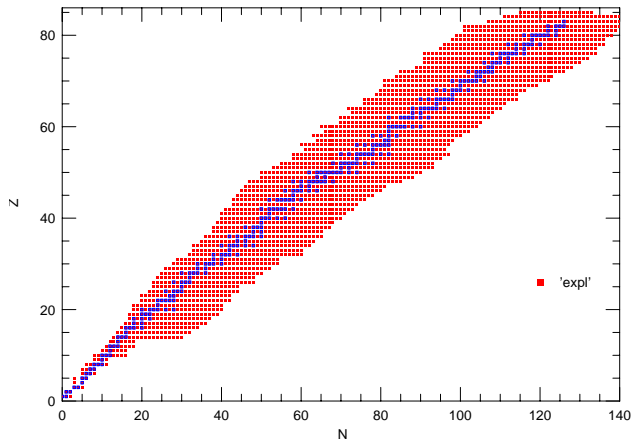}
\end{center}
\caption{Reaction network for explosive burning.  \label{fig:net}}
\end{figure}

The nuclear reaction network during the explosive
phase (as the most extreme case in our calculation)
contains about 2350 isotopes and is shown in Fig.\ \ref{fig:net}. It is evident
that there are many isotopes far off stability included for which experimental
information is scarce. This is even more true in more exotic scenarios
such as the r-- and rp--processes which involve isotopes close to the neutron--
and proton--dripline, respectively. (In our calculations we do not follow the
proposed r--process in the $\nu$--wind emanating from the proto--neutron
star shortly after the collapse of the Fe core.)

Important are weak reactions and nuclear reactions with nucleons and $\alpha$
particles. Decay data is also available further off stability whereas
nuclear reaction cross sections involving nucleons and light ions are
practically known only for stable nuclei. The majority of the latter reactions
can be described in the framework of the statistical model (Hauser-Feshbach
theory) which describes the reaction proceeding via the formation of a
compound nucleus and averages over resonances\cite{rtk97}. 
Many nuclear properties enter the computation of the HF cross sections:
mass differences (separation energies), optical potentials, Giant Dipole
Resonance widths,
level densities. The resulting transmission coefficients can be modified due to
pre-equilibrium effects which are included in width fluctuation
corrections (see also a previous paper\cite{rtk97} and references therein)
and by isospin effects. It is in the description of the nuclear
properties where the various HF models differ. In astrophysical applications 
usually different aspects are emphasized
than in pure nuclear physics investigations. Many of
the latter in this long and well established field were focused on
specific reactions, where all or most "ingredients"
were deduced from
experiments. As long as the reaction mechanism is identified properly,
this will produce highly accurate cross sections.
For the majority of nuclei in astrophysical applications such
information is not available. The real challenge is thus not the
application of well-established models, but rather to provide all the
necessary ingredients
in as reliable a way as possible, also for nuclei where no such
information is available.

\subsection{Statistical Model Rates}
As the basis for the creation of our reaction rate set we used
statistical model calculations obtained with the NON-SMOKER 
code\cite{rtk97,rt00}. A library of theoretical reaction rates calculated with
this code and fitted to an analytical function --- ready to be incorporated
into stellar model codes --- was published recently\cite{rt00}. It includes
rates for all possible targets from proton-- to neutron--dripline and between
Ne and Bi, thus being the most extensive published library of theoretical 
reaction rates to date. It also offers rate sets for a number of mass models
which are suited for different purposes. For the network described here we
utilized the rates based on the FRDM set.

\subsection{$\alpha$ Particles: Isospin Effects}
The consideration of isospin effects has two major effects on statistical cross
sections in astrophysics\cite{r98}:
the suppression of $\gamma$ widths for reactions involving
self-conjugate nuclei and the suppression of the neutron emission
in proton-induced reactions. Here, we only discuss the former.
In the case of ($\alpha$,$\gamma$) reactions on targets with $N=Z$, the
cross sections will be heavily suppressed because $T=1$ states cannot be
populated due to isospin conservation. A suppression will also be found
for capture reactions leading into self-conjugate nuclei, although
somewhat less pronounced because $T=1$ states can be populated according
to the isospin coupling coefficients.
In previous reaction rate calculations\cite{woo78,cow91} the
suppression of the $\gamma$--widths was treated completely
phenomenologically by employing arbitrary and mass-independent
suppression factors.
In the NON-SMOKER code, the
appropriate $\gamma$ widths are automatically obtained, by explicitly
accounting for $T^<$ and $T^>$ states\cite{rtgw00}.
The astrophysical importance of $\alpha$ capture on target nuclei
with $N=Z$ is manifold. In the Ne- and O-burning phase of massive
stars, alpha capture reaction sequences are initiated at $^{24}$Mg
and $^{28}$Si, respectively, and determine the abundance
distribution prior to the Si-burning phase.
Nucleosynthesis in explosive Ne and explosive O burning in type II
supernovae depend on reaction rates for $\alpha$ capture on
$^{20}$Ne to $^{36}$Ar.  An $\alpha$
capture chain on such self-conjugate nuclei actually determines
the production of $^{44}$Ti \cite{hoff98}, 
which contributes to
the light curve by its $\beta$ decay to $^{44}$Ca via $^{44}$Sc.

\subsection{$\alpha$ Particles: Optical $\alpha$+Nucleus Potentials}
A further complication in the treatment of reactions on intermediate and
light nuclei involving $\alpha$
particles is the limited success in defining an appropriate optical potential,
especially for the low energies typical for astrophysical environments.
Early astrophysical studies (e.g.\cite{woo78}) made use of
simplified equivalent square well potentials and the black nucleus
approximation. It is equivalent to a fully absorptive potential, once a
particle has entered
the potential well and therefore does not permit resonance effects.
This leads to deviations from experimental data at low energies,
especially in mass regions where broad resonances in the continuum can
be populated\cite{hoff98}. An additional effect, which is only
pronounced for $\alpha$ particles, is that absorption in the Coulomb
barrier\cite{hoff98} is neglected in this approach.
Improved calculations have to employ appropriate {\it global} optical
potentials which make use of imaginary parts describing the absorption.
In the case of $\alpha$-nucleus potentials, there were only very few 
global parametrizations attempted at astrophysical energies, also due to
the scarcity of experimental data in the energy region of interest.
The high Coloumb barrier makes a
direct experimental approach very difficult at low energies.
Current astrophysical calculations mostly
employ a phenomenological Saxon--Woods
potential based on extensive data\cite{mcf66}.
Future improved $\alpha$ potentials have
to take into account the mass- and energy-dependence of the potential.
Extended investigations of $\alpha$ scattering data\cite{mohr94,atz96}
have shown that the data can best be described with folding
potentials.
Few attempts\cite{r98} have been made to construct such an improved global potential.
Nevertheless, the postulation of such an optical
potential close to or below the Coulomb barrier remains one of the major
challenges. More experimental data is clearly needed.
\begin{figure}[t]
\begin{center}
\includegraphics*[bb=50 50 792 550,height=16pc]{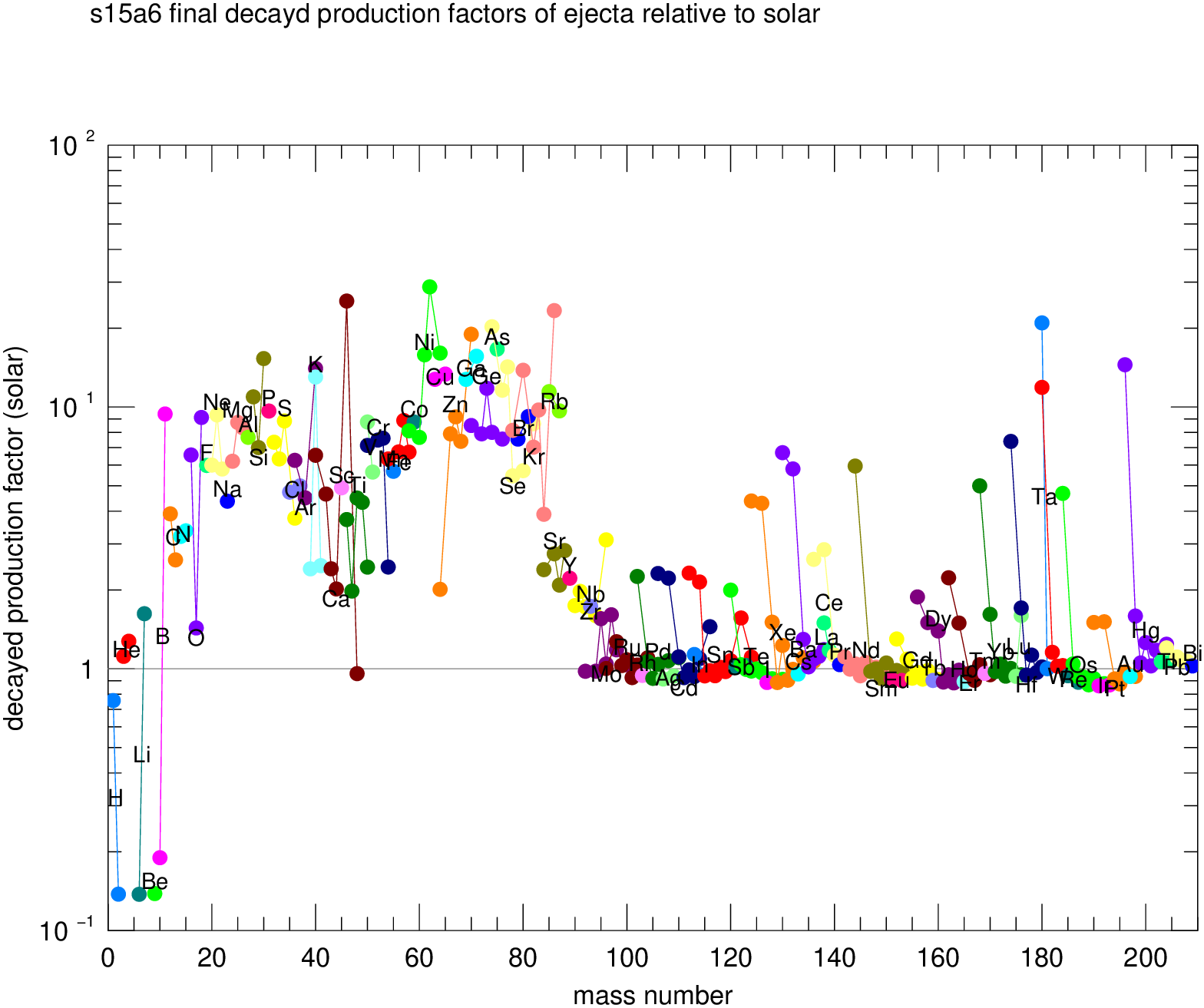}
\end{center}
\caption{Production factors in the ejecta of a 15 $M_{\odot}$ star relative to
solar abundance.  \label{fig:yields}}
\end{figure}

\section{Further Inputs}
Further nuclear input were updated rates from experimental cross sections
for light as well as heavy nuclei and updated beta-decay rates. New
predictions of weak rates\cite{lamar00} were also included. In respect to
earlier simulations\cite{ww95} we also used updated neutrino loss rates and
opacity tables (OPAL95), and
consider mass loss due to stellar wind. For further details, refer to our
other papers\cite{hhrw00,rhhw00}.
\section{Results and Summary}
For the first time, we studied consistently the production of all isotopes up 
to Bi during the pre-supernova evolution and the type II supernova
explosion of a massive star. 
Exemplary for our results, 
the production factors
of a 15 $M_{\odot}$ star are shown in Fig.\ \ref{fig:yields}.
The revised weak rates introduce an important change mainly during core
silicon burning and thereafter leading to an increase of the
central $Y_e$ and smaller Fe core masses at the onset of core collapse.
In addition to the well-known strong dependence of the stellar structure on
the $^{12}$C($\alpha$,$\gamma$)$^{16}O$ rate, we also found the 
($\alpha$,n)--($\alpha$,$\gamma$) branching on $^{22}$Ne to be an
important candidate for further laboratory study. It sensitively determines the
strength of the s--process in the SN models.

Summarizing, the progress in predictions of nuclear reactions has made it
possible to consistently study the nucleosynthesis in a type II supernova
model over a wide range of nuclear masses.
The new investigations
also underline the importance of new experimental and theoretical studies
of specific nuclear properties.

\section*{Acknowledgments}
This research was supported, in part, by DOE (W-7405-ENG-48), NSF
(AST 97-31569, INT-9726315), and the Alexander von
Humboldt Foundation (FLF-1065004). T. R. acknowledges support by a PROFIL
professorship from the Swiss NSF (grant 2124-055832.98).


\begin{thebibliography}{99}
\bibitem{hhrw00}A. Heger, R.D. Hoffman, T. Rauscher, S.E. Woosley, in {\em
Proc.\ X Workshop on Nuclear Astrophysics}, eds.\ W. Hillebrandt, E. M\"uller
(MPA, Garching 2000), in press. (astro-ph/abs/0006350)
\bibitem{rhhw00}T. Rauscher, R.D. Hoffman, A. Heger, S.E. Woosley,
\Journal{\em Ap.\ J.}{}{in prep.}{2000}.
\bibitem{rtk97}T. Rauscher, F.-K. Thielemann, K.-L. Kratz, \Journal{\em
Phys.\ Rev. C}{56}{1613}{1997}.
\bibitem{rt00}T. Rauscher, F.-K. Thielemann, \Journal{\em At. Data 
Nucl.\ Data Tables}{75}{1}{2000}.
\bibitem{rtgw00}T. Rauscher, F.-K. Thielemann, J. G\"orres, M.C. Wiescher,
\Journal{\em Nucl.\ Phys.}{A675}{695}{2000}.
\bibitem{r98}T. Rauscher, in {\em Proc.\ Symp.\ "Nuclei in the Cosmos V"},
eds.\ N. Prantzos, S. Harissopoulos (Editions Fronti\`eres, Gif-sur-Yvette
1998), p.\ 484.
\bibitem{woo78}S.E. Woosley, W.A. Fowler, J.A. Holmes, B.A. Zimmerman,
\Journal{\em At. Data Nucl.\ Data Tables}{18}{306}{1978}.
\bibitem{cow91}J.J. Cowan, F.-K. Thielemann, J.W. Truran,
\Journal{\em  Phys.\ Rep.}{208}{267}{1991}.
\bibitem{hoff98}R.D. Hoffman {\it et al}, \Journal{\em Ap.\ J.}{521}{735}{1999}.
\bibitem{mcf66}L. McFadden, G.R. Satchler, \Journal{\em Nucl.\ Phys.}{84}{177}{1966}.
\bibitem{mohr94}P. Mohr {\it et al}, \Journal{\em Phys.\ Rev.\ C}{55}{1523}{1997}.
\bibitem{atz96}P. Mohr, \Journal{\em Phys.\ Rev.\ C}{61}{045802}{2000}.
\bibitem{lamar00}K. Langanke, G. Mart\'inez-Pinedo, \Journal{\em Nucl.\ Phys.}{A673}{481}{2000}.
\bibitem{ww95}S.E. Woosley, T.A. Weaver, \Journal{\em Ap.\ J. Suppl.}{101}{181}{1995}.
\bibitem{awml00}A. Heger, S.E. Woosley, G. Mart\'inez-Pinedo, K.
Langanke, \Journal{\em Ap.\ J.}{}{in prep.}{2000}.

\end{thebibliography}
\end{document}